\documentclass[12pt]{iopart}
\usepackage[utf8]{inputenc}
\usepackage[english]{babel}
\usepackage{iopams} 
\usepackage[pdftex]{hyperref} 

 \addto\extrasenglish{%
  }

\newcommand{\circled}[1]{\tikz[baseline=(char.base)]{
            \node[shape=circle,draw,inner sep=2pt] (char) {#1};}}

\expandafter\let\csname equation*\endcsname\relax
\expandafter\let\csname endequation*\endcsname\relax
\usepackage{amsmath}
\usepackage{amssymb}
\usepackage{color}
\usepackage{geometry}

\usepackage[pdftex]{graphicx}
\hypersetup{colorlinks=true,
	linkcolor=blue,
	citecolor=blue,
	urlcolor=blue,
	pdfauthor={Ansh Patel IPP, Anshkumarhimanshu.Patel@ipp.mpg.de},
	pdfsubject={Patel 2024 Modelling of shattered pellet injection experiments on the ASDEX Upgrade tokamak},
	pdftitle={Modelling of shattered pellet injection experiments on the ASDEX Upgrade tokamak}
}
\DeclareGraphicsExtensions{.pdf,.png}

\usepackage{float}
\usepackage{caption}
\usepackage[labelformat=empty]{subcaption}
\usepackage{siunitx}
\usepackage{todonotes}
\usepackage{tikz}
\usetikzlibrary{positioning,fit}

\usepackage{orcidlink}

\usepackage[numbers,sort&compress]{natbib}
\usepackage{hypernat}

\newcommand{\backavg}{\beta_d}
\begin{document}

\title[Modelling of SPI on AUG]{Modelling of shattered pellet injection experiments on the ASDEX Upgrade tokamak}
\author{
Ansh Patel$^1$\orcidlink{0000-0002-7349-3243},
A.~Matsuyama$^2$\orcidlink{0000-0002-6634-2025},
G.~Papp$^1$\orcidlink{0000-0003-0694-5446},
M.~Lehnen$^3$\footnote{Deceased},
J.~Artola$^3$,
S.~Jachmich$^3$,
E.~Fable$^1$\orcidlink{0000-0001-5019-9685},
A.~Bock$^1$,
B.~Kurzan$^1$,
M.~Hoelzl$^1$\orcidlink{0000-0001-7921-9176},
W.~Tang$^1$,
M.~Dunne$^1$,
R.~Fischer$^1$\orcidlink{0009-0000-6205-4731},
P.~Heinrich$^1$\orcidlink{0000-0003-1823-5257}, 
the ASDEX Upgrade Team$^a$
 and the EUROfusion Tokamak Exploitation Team$^{b}$}
\address{$^1$Max Planck Institute for Plasma Physics, Boltzmannstr. 2, D-85748 Garching, Germany}
\address{$^2$Graduate School of Energy Science, Kyoto University, Uji, Kyoto
611-0011, Japan}
\address{$^3$ITER Organization, Route de Vinon-sur-Verdon, CS 90 046 13067 St.~Paul-lez-Durance, France}
\address{$^a$See the author list of \href{https://doi.org/10.1088/1741-4326/ad249d}{H.~Zohm~\etal 2024 Nucl. Fusion}} 
\address{$^b$See the author list of \href{https://doi.org/10.1088/1741-4326/ad2be4}{E.~Joffrin~\etal 2024 Nucl. Fusion}}
\ead{Anshkumarhimanshu.Patel@ipp.mpg.de}

\begin{abstract}
 In a shattered pellet injection (SPI) system the penetration and assimilation of the injected material depends on the speed and size distribution of the SPI fragments.
ASDEX Upgrade (AUG) was recently equipped with a flexible SPI to study the effect of these parameters on disruption mitigation efficiency. 
In this paper we study the impact of different parameters on SPI assimilation with the 1.5D INDEX code.
Scans of fragment sizes, speeds and different pellet compositions are carried out for single SPI into AUG H-mode plasmas. 
We use a semi-empirical thermal quench (TQ) onset condition to study the material assimilation trends. 
For mixed deuterium-neon pellets, smaller/faster fragments start to assimilate quicker. 
However, at the expected onset of the global reconnection event (GRE), larger/faster fragments end up assimilating more material. 
Variations in the injected neon content lead to a large difference in the assimilated neon for neon content below $<10^{21}$ atoms. For larger injected neon content, a self-regulating mechanism limits the variation in the amount of assimilated neon. 
We use a back-averaging model to simulate the plasmoid drift during pure deuterium injections with the back-averaging parameter determined by a interpretative simulation of an experimental pure deuterium injection discharge. 
Again, larger and faster fragments are found to lead to higher assimilation with the material assimilation limited to the plasma edge in general, due to the plasmoid drift. 
The trends of assimilation for varying fragment sizes, speeds and pellet composition qualitatively agree with the previously reported experimental observations. 
\end{abstract}

\submitto{\NF}
\vspace*{-1in}  
\maketitle

\section{Introduction}
A disruption mitigation system (DMS) is necessary for reactor-relevant tokamaks like ITER~\cite{lehnen_disruptions_2015} in order to ensure the preservation of machine components throughout their designated operational lifespan. 
To address the intense heat and electromagnetic loads that can occur during a disruption, a shattered pellet injection (SPI) system will be utilized by ITER~\cite{lehnen_rd_2018, luce_progress_2021, jachmich_shattered_2022}. 
An SPI system injects material into the plasma in the form of a cryogenic pellet that is shattered on a shatter unit, e.g. a bend in the guide tube before entering the plasma.
The shattering allows more efficient material assimilation due to the increased surface area exposure to the background plasma \cite{lehnen_rd_2018}. 
The ITER DMS is required to reduce the vessel forces by a factor of 2-3 and heat loads by at least 90\%~\cite{lehnen_disruptions_2015, hollmann_status_2014} when operating at its highest currents and energy. 
To aid the design, optimisation and commissioning of the ITER DMS, experiments are being carried out in present tokamaks such as DIII-D~\cite{herfindal_injection_2019}, JET~\cite{s_jachmich_shattered_2023}, KSTAR~\cite{park_experimental_2021}, J-TEXT~\cite{li_comparison_2021}, HL-2A~\cite{xu_preliminary_2020}, EAST~\cite{yuan_first_2023} and ASDEX Upgrade (AUG)~\cite{dibon_design_2023,Heinrich2024Recipes}.
An important parameter of an SPI system which affects material penetration and assimilation is the size and speed of the fragments generated by the shattered pellets. 
Here, AUG plays an important role due to its highly flexible SPI system capable of varying fragment sizes and speeds independently. 
This capability was enabled by its triple barrel injectors ~\cite{dibon_design_2023}, where each independent guide tube can be equipped with different shatter heads. 
Experiments carried out in the 2022 experimental campaign~\cite{heinrich_analysis_2024,Heinrich2024Radiated} utilised three different shattering tubes with different cross sections and shattering angles. 

To make predictions for the ITER DMS, accurate modelling of the plasma response to an SPI system and validation with present experiments is required. 
For this purpose, 3D MHD codes have proven successful in studying the material assimilation, MHD activity and associated radiation characteristics~\cite{hu_3d_2018, hu_radiation_2021, kim_shattered_2019, nardon_fast_2020, hu_plasmoid_2024, kong_interpretative_2024}. 
However, these simulations are computationally expensive, limiting our ability to carry out very detailed parameter studies. For this purpose, reduced models are utilised to carry out larger parametric scans to understand the role of key SPI parameters and validate their relationship with material assimilation.

In this paper, we use the 1.5D disruption simulator code INDEX~\cite{matsuyama_transport_2022} to validate the trends in penetration and material assimilation in the AUG tokamak with three key SPI parameters: fragment sizes, speeds, and pellet composition. 
We consider pure deuterium and mixed deuterium-neon pellets at various mixing ratios.
A back-averaging model~\cite{jardin_fast_2000} is used to simulate the plasmoid drift that is experienced by pure deuterium pellets. 
The simulation results are compared with trends of assimilation from the experimental analysis.

The paper is structured as follows: \autoref{sec:INDEX_AUG_intro} briefly describes the INDEX code and the AUG-specific inputs used in the simulations, along with a detailed  description of the plasma response for a mixed deuterium-neon pellet. 
The results of the parametric simulation scans are discussed in \autoref{sec:parametric_scans} and are categorised by the pellet composition, and further sub-categorised into scans of fragment sizes and speeds. 
Experimental comparisons, wherever available, are also presented. 

\section{Modelling AUG SPI using INDEX \label{sec:INDEX_AUG_intro}}
INDEX is an axisymmetric simulation code~\cite{matsuyama_transport_2022} that self-consistently solves 1D transport equations in the magnetic flux coordinates along with a 2D Grad-Shafranov equilibrium solver~\cite{hinton_theory_1976} with the boundary conditions given by solving a series of circuit equations of toroidally continuous filaments. 
The free-boundary equilibrium is evolved in time with iterations to a 1D magnetic diffusion equation. 
The densities of the various ion species are individually modelled, accounting for different charge states, while assuming  temperature equilibration between the charge states. 
Ionisation and recombination rates are sourced from the OpenADAS database~\cite{summers_ionization_2006}. 
For the SPI module, INDEX models the shattered fragments as spheres whose size distribution is obtained from a statistical fragmentation model~\cite{hu_3d_2018,parks_modeling_2016}. 
The fragment velocities are sampled from a Gaussian distribution centered at the pellet velocity with a user-defined velocity spread. 
The 3D location of the fragments are mapped to the 1D flux coordinates at each time step. 
The ablation from the shattered fragments is estimated using the Neutral Gas Shielding (NGS) model~\cite{parks_theoretical_2017}. 
During a simulation, INDEX traces the motion of multiple pellet markers with given fragment velocities, calculates the recession in fragment sizes using the NGS model and assumes the ablated particles are deposited as a flux-surface-averaged neutral particle source. 
Further details on the employed equations and details can be found in the paper by Matsuyama~\cite{matsuyama_transport_2022}. 

\subsection{Simulation parameters}
\begin{figure}[htb]
\centering
\includegraphics[width = 0.5\textwidth]{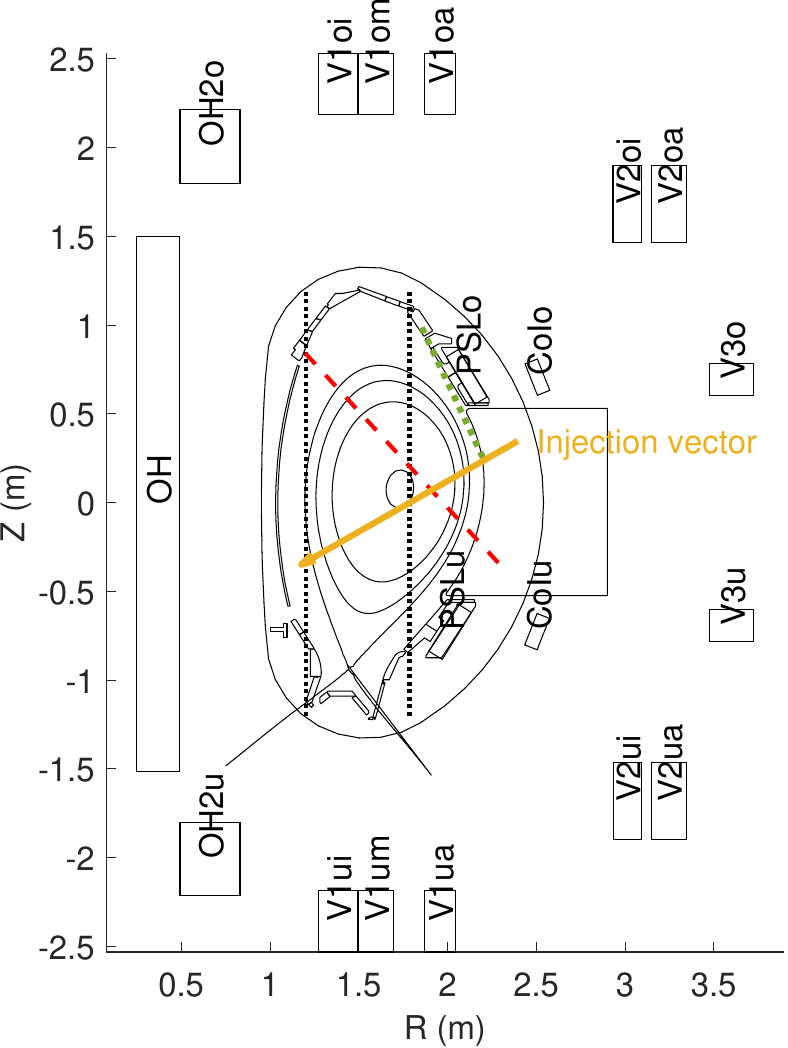}
\caption{Poloidal cross-section showing coil and vessel elements, along with equilibrium contours of the $q=1,2,3$ surfaces, the separatrix; and the SPI injection vector (orange arrow). Green dotted line marks an edge AXUV line of sight (LOS) intersecting the entry location of the fragments. Red dashed line marks a core SXR LOS. Vertical black dotted lines mark the \si{CO_2} interferometer LOS. }
\label{fig:AUG_INDEX_inputs_poloidal_cross_section} 
\end{figure}

For the free-boundary equilibrium solver, the poloidal location of the Ohmic and poloidal field coils have been adopted from the AUG flight simulator~\cite{fable_modeling_2022} and are shown in \autoref{fig:AUG_INDEX_inputs_poloidal_cross_section}.  
For the sake of simplicity, only a single injection vector is utilised for all the simulations corresponding to the $25^\circ$ rect. shatter head geometry (shown in \autoref{fig:AUG_INDEX_inputs_poloidal_cross_section}). The largest impact of the various shattering angles comes via the change of the fragment distribution, with the different injection direction having a less significant role~\cite{Hoelzl2020First}.

\begin{figure}[htb]
   \centering
     \begin{subfigure}[t]{0.49\textwidth}
         \begin{tikzpicture}                 
         \node (inner sep=0pt) (image) at (0,0)  {\includegraphics[width=\textwidth]{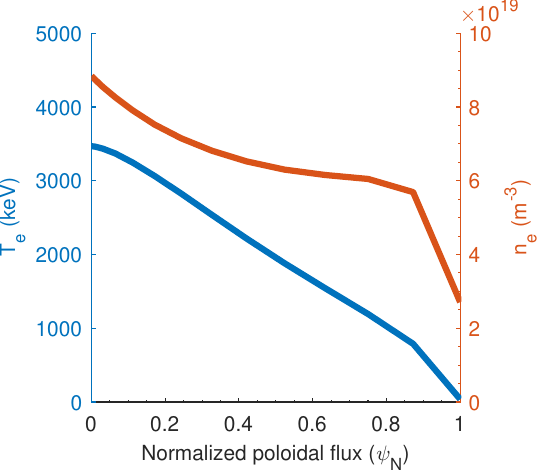}};
        \node[overlay, below right = 0.7cm and 4cm of image] at (image.north west) {(a)};
         \end{tikzpicture}
     \end{subfigure}
     \begin{subfigure}[t]{0.49\textwidth}
          \begin{tikzpicture}                 
         \node (inner sep=0pt) (image) at (0,0)  {\includegraphics[width=\textwidth]{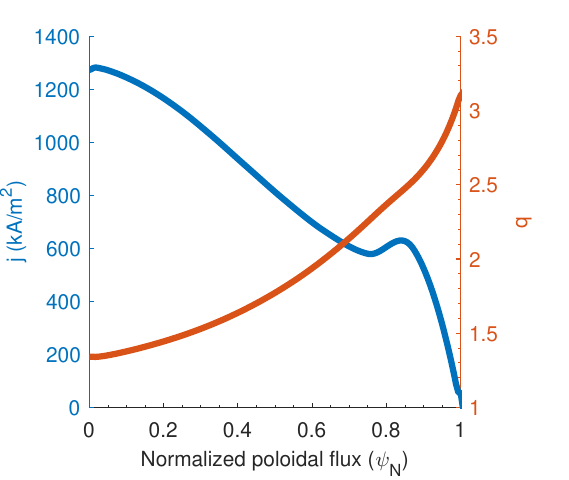}};
        \node[overlay, below right = 0.7cm and 4cm of image] at (image.north west) {(b)};
        \end{tikzpicture}
        \label{fig:INDEX_inputs_J_q}          
     \end{subfigure}
     \vspace{-2em}
\caption{Radial plasma profiles used for initial plasma equilibrium convergence, (a) Electron temperature (left axis) and density (right axis); (b) Current density (left axis) and q-profile (right axis)}
\label{fig:INDEX_input_radial_profiles}
\end{figure}

To initialise the simulation, a set of radial profiles of electron, temperature and current density are required corresponding to the plasma equilibrium. 
The inputs used in the simulations presented in this paper are based on the AUG discharge \#40355 \cite{Heinrich2024Radiated}. 
The density and temperature profiles were obtained from the Integrated Data Analysis (IDA) diagnostic~\cite{fischer_integrated_2010} which combined measurements from lithium beam emission spectroscopy (LIB), deuterium cyanide (DCN) laser interferometry, electron cyclotron emission (ECE), and Thomson scattering (TS) spectroscopy to determine electron density and temperature. 
The temperature profile was further modified to match the pressure profile from the CLISTE plasma equilibrium \cite{mc_carthy_analytical_1999} to take into account the contribution from fast particles not included in the IDA thermal profiles. 
This adjustment serves as a practical approach to partially account for the ablation from fast particles, which are not included in the IDA thermal profiles.
The same profiles have also been used for AUG JOREK simulations \cite{WTang_non_linear_AUG_2024}. 
The reconstructed equilibrium and measured electron density and current density profiles and are shown in \autoref{fig:INDEX_input_radial_profiles}. 
Assuming $T_e = T_i$, the thermal energy content before the SPI is 564~kJ. 
Similar to previous ITER SPI simulations with INDEX~\cite{matsuyama_transport_2022}, the particle transport and heat diffusion coefficient were set to $D_\alpha = 2 \si{m^2 /s}$ and $K_\alpha = 4.5 \si{m^2 /s}$ respectively for the entire simulation and were assumed to be radially uniform.  
For all the simulations in this paper, the velocity dispersion is set to $\Delta v/v = \pm 40\%$ and the vertex angle of the injection cone, i.e. the spatial toroidal and the poloidal spread is set to $\pm 20^\circ$~\cite{Gebhart2020Experimental}. 

\subsection{Characteristic plasma response during impurity SPI \label{ssec:characteristic_response}}

To assist the interpretation of parametric scans reported in this paper, first we describe the plasma response to a 200~\si{m/s} mixed deuterium-neon injection with 10\% neon molar fraction. 
The pellet diameter is set to 8~\si{mm} and the pellet length to 10~\si{mm}. 
The simulation is initialised at the time of shattering, where a distribution of fragments is generated for a user-defined number of fragments, in this case, $N_\text{frags} = 200$.

\begin{figure}[htb]
\begin{minipage}[t]{0.45\textwidth}
  \setlength{\belowcaptionskip}{-1cm}  
  \setlength{\abovecaptionskip}{\baselineskip}
   \centering        
     \begin{subfigure}[t]{\textwidth}
         \centering
         \begin{tikzpicture}
            \node (inner sep=0pt) (image) at (0,0) {\includegraphics[width=\textwidth]{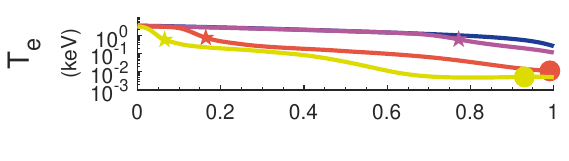}};    

            \node[draw, circle, minimum size=0.8em, font=\fontsize{10}{10}\selectfont, inner sep=1 pt, above left = 1.3cm and 2.8cm of image] at (image.south east) {4};

            \node[draw, circle, minimum size=0.8em, font=\fontsize{10}{10}\selectfont, inner sep=1 pt, above left = 1cm and 1.6cm of image] at (image.south east) {5};

            \node[overlay, below right = 0.6cm and 1.75cm of image] at (image.north west) {(a)};
         \end{tikzpicture}                  
         \caption{}
         \label{fig:example_sim_subplots_a}
     \end{subfigure}    
    \vspace{-1em}   
    \begin{subfigure}[t]{\textwidth}
         \centering
         \begin{tikzpicture}
            \node (inner sep=0pt) (image) at (0,0) {\includegraphics[width=\textwidth]{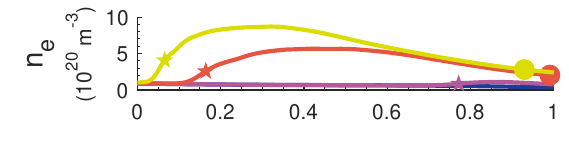}};    

            \node[draw, circle, minimum size=0.8em, font=\fontsize{10}{10}\selectfont, inner sep=1 pt, above left = 1.2cm and 3.4cm of image] at (image.south east) {3};

            \node[overlay, below right = 0.4cm and 2cm of image] at (image.north west) {(b)};
         \end{tikzpicture}        
         \caption{}
         \label{fig:example_sim_subplots_b}
     \end{subfigure}
 \vspace{-1em}  
    \begin{subfigure}[t]{\textwidth}
         \centering
         \begin{tikzpicture}
            \node (inner sep=0pt) (image) at (0,0) {\includegraphics[width=\textwidth]{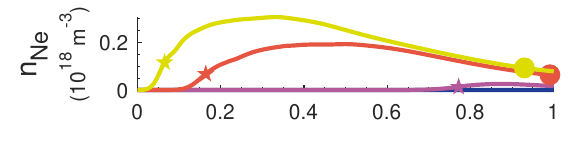}};    

            \node[overlay, below right = 0.4cm and 2cm of image] at (image.north west) {(c)};
         \end{tikzpicture}            	 
         \caption{}
         \label{fig:example_sim_subplots_c}
     \end{subfigure}
     \vspace{-1em}  
     \begin{subfigure}[t]{\textwidth}
         \centering
         \begin{tikzpicture}
            \node (inner sep=0pt) (image) at (0,0) {\includegraphics[width=\textwidth]{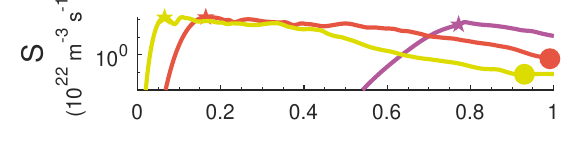}};    

            \node[draw, circle, minimum size=0.8em, font=\fontsize{10}{10}\selectfont, inner sep=1 pt, below left = 0cm and 0.3cm of image] at (image.north east) {1};

            \node[draw, circle, minimum size=0.8em, font=\fontsize{10}{10}\selectfont, inner sep=1 pt, below right = -0.1cm and 1.95cm of image] at (image.north west) {7};

            \node[overlay, below right = 0.6cm and 2.1cm of image] at (image.north west) {(d)};
            
         \end{tikzpicture}          
         \caption{}
         \label{fig:example_sim_subplots_d}
     \end{subfigure}
     \vspace{-1em}  
     \begin{subfigure}[t]{\textwidth}
         \centering
         \begin{tikzpicture}
            \node (inner sep=0pt) (image) at (0,0) {\includegraphics[width=\textwidth]{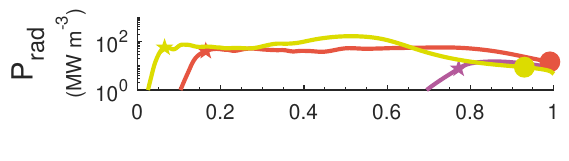}};    

            \node[draw, circle, minimum size=0.8em, font=\fontsize{10}{10}\selectfont, inner sep=1 pt, above left = 0.7cm and 1.7cm of image] at (image.south east) {2};

            \node[overlay, below right = 0.1cm and 2.3cm of image] at (image.north west) {(e)};
            
         \end{tikzpicture}            	 
         \caption{}
         \label{fig:example_sim_subplots_e}
     \end{subfigure}
     \vspace{-1em}  
     \begin{subfigure}[t]{\textwidth}
         \centering
         \begin{tikzpicture}
            \node (inner sep=0pt) (image) at (0,0) {\includegraphics[width=\textwidth]{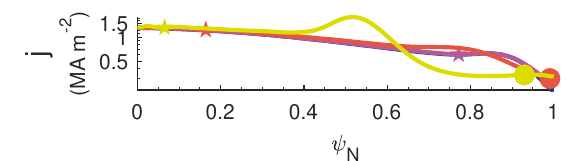}};    
            \node[overlay, below right = 0.7cm and 2cm of image] at (image.north west) {(f)};            

            \node[draw, circle, minimum size=0.8em, font=\fontsize{10}{10}\selectfont, inner sep=1 pt, above left = 1.8cm and 2.2cm of image] at (image.south east) {6};
            
            \node[overlay, anchor = north west, below right = -0.9cm and 2cm of image.north west] (overlap) {\includegraphics[width = 1.6cm]{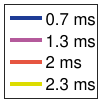}};
         \end{tikzpicture}            	 
         \caption{}
         \label{fig:example_sim_subplots_f}
     \end{subfigure}        
\setlength{\belowcaptionskip}{\baselineskip}
\caption{Profiles for various SPI relevant plasma parameters during an injection plotted against normalized poloidal flux: (a) electron temperature, (b) electron density, (c) neon density, (d) neutral particle source function~\cite{matsuyama_transport_2022}, (e) radiated power, (f) current density. The circled numbers indicate key plasma dynamics which are discussed in the text.} 
\label{fig:example_sim_subplots}
\end{minipage}
\hspace{0.05\textwidth} 
\begin{minipage}[t]{0.45\textwidth}
    \centering
    \vspace{1cm}     
        \includegraphics[width=\textwidth]{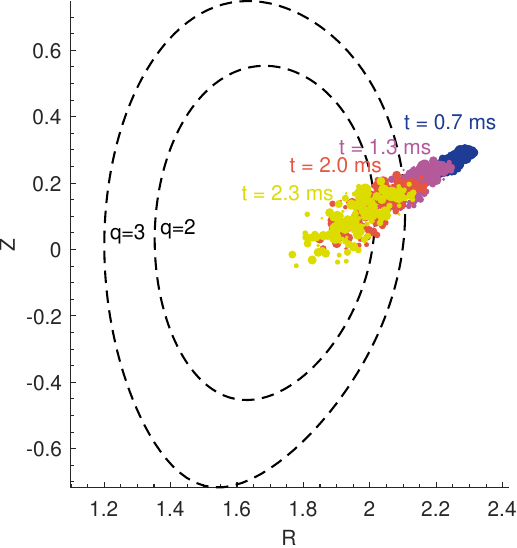}
        \caption{Fragment locations at the times of plotted profiles in \autoref{fig:example_sim_subplots} (point size enlarged, not representative of fragment sizes).}
        \label{fig:example_sim_frags_location}
\end{minipage}
\end{figure}
\clearpage

The profiles of relevant plasma parameters during the SPI are shown in \autoref{fig:example_sim_subplots}. 
Two key fragment location parameters are shown in the form of pentagram and circular scatter points: the front of the fragment plume corresponding to the location of the fragment with 99\% speed percentile and rear of the plume corresponding to the fragment with 25\% speed percentile, respectively. 
Additionally, the fragment locations at the time of plotted profiles are shown in \autoref{fig:example_sim_frags_location}.
As the fragments enter the plasma, at 0.7~ms, an initial rise in the neutral particle source function is observed at the plasma edge, as marked in \autoref{fig:example_sim_subplots_d} as \circled{1} and a slight increase in radiated power \circled{2} is observed corresponding to line radiation from the injected neon particles. 
Eventually, as the fragments travel further in the plasma, additional neutral particles are introduced towards the core, the electron density starts to increase (\autoref{fig:example_sim_subplots_b}, \circled{3}) and the electron temperature (\autoref{fig:example_sim_subplots_a}, \circled{4}) decreases. 
Note that the ablation front follows the front of the plume and the radiation front lags only slightly behind. 
Finally, a radiative cold front ($<10 \si{eV}$) starts to form at the plasma edge (\autoref{fig:example_sim_subplots_a}, \circled{5}) at 2.3~ms, leading to strong gradients in the current density (\autoref{fig:example_sim_subplots_f}, \circled{6}). 
Since the plasma edge is cold and the bulk of the fragments are towards the plasma core which is still relatively hot, more neutrals are introduced in the plasma core (\autoref{fig:example_sim_subplots_d}, \circled{7}).

Like in Massive Gas Injection (MGI) experiments~\cite{hollmann_measurements_2005, granetz_gas_2007, bozhenkov_generation_2008, pautasso_generation_2020, reux_experimental_2010}, once the cold front crosses the $q=2$ surface, corresponding to $\psi_N \approx 0.8$ in \autoref{fig:example_sim_subplots}, strong tearing modes are expected to be destabilised and initiate a thermal quench. 
Although the cold front lags behind in time compared to the movement of the ablation and the radiation front however it can still lead to the onset of the global reconnection event (GRE) before majority of the material is assimilated. 
In this simulation, only $\approx 9\%$ of the injected material is ablated inside the plasma volume at the GRE onset.
The time between when the fragments cross the AXUV LOS shown in \autoref{fig:AUG_INDEX_inputs_poloidal_cross_section} and the cold front ($<10 \si{eV}$) reaches the $q=2$ surface is referred to as the simulated pre-GRE duration. 
In the present simulation, the pre-GRE duration is estimated as 1.25~ms.  
Modelling the plasma dynamics in the GRE phase and beyond is outside the scope of this article and will be considered in future work. 
Hence, while the presented simulations might extend beyond the expected GRE onset, no changes in the transport parameters are made. 
Parallel transport losses are not included in the model and might contribute to thermal energy loss especially in the presence of stochastic fields.  

\section{Parametric scans \label{sec:parametric_scans}}
\subsection{Deuterium-neon mixture injections}
\subsubsection{Fragment sizes \label{sssec:frag_size_var_D_Ne}}

\begin{figure}[htb]
\centering
\begin{tikzpicture}
\node (inner sep=0pt) (image) at (0,0) {\includegraphics[width=0.5\linewidth]{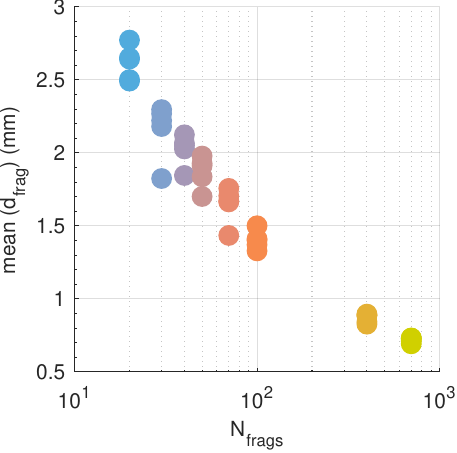}};
  \node[overlay, anchor = south west, below left = 0.6cm and 0.5cm of image.north east] (overlap) {\includegraphics[width = 2cm]{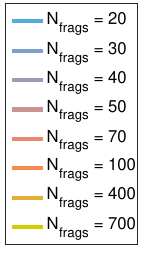}};
\end{tikzpicture}
\caption{Relation between sampled number of fragments and mean fragment size.}
\label{fig:frag_size_var_relation_btwn_num_and_size} 
\end{figure}

In this subsection, we discuss a set of simulations of varying fragment sizes for 10\% neon and 90\% deuterium SPI molecules. 
The plasmoid drift is not taken into account for any of the mixed deuterium-neon injections presented in this subsection and is only considered for pure deuterium injections discussed in \autoref{ssec:pure_deuterium_pellets}.
The fragment sizes were varied indirectly by sampling different number of fragments from a fixed pellet size of 8~mm diameter and 10~mm length. 
The mean fragment velocity was set to 230~\si{m/s} for all cases along with a fixed velocity dispersion of $\Delta v/v = 40\%$. 
The resulting relation between number of sampled fragments and mean fragment size is shown in \autoref {fig:frag_size_var_relation_btwn_num_and_size}. 
Multiple realisations were sampled from the statistical size distribution to ensure that general trends are observed. 
Different realisations of the size distribution for the same number of total fragments are plotted with the same colour scatter points in \autoref{fig:frag_size_var_relation_btwn_num_and_size} and for the rest of the paper as well.

\begin{figure}[htb]
    \centering
    \begin{subfigure}[t]{0.5\textwidth}
         \centering
         \begin{tikzpicture}
         \node (inner sep=0pt) (fig1) at (0,0) {\includegraphics[width=\textwidth]{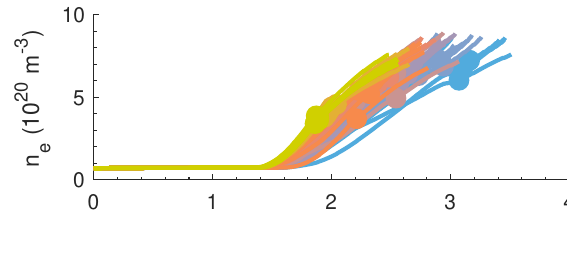}}; 
         \node[overlay, below right=0.6cm and 4cm of fig1.north west] (fig1) {(a)};
         \end{tikzpicture}             
    \end{subfigure}

    \begin{subfigure}[t]{0.5\textwidth}
    \centering
    \begin{tikzpicture}

    \node (inner sep=0pt) (fig2) at (0,0) {\includegraphics[width=\textwidth]{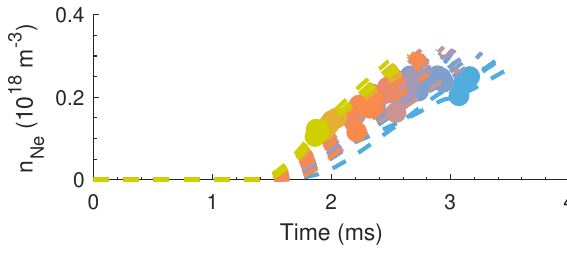}};  
    
    \node[overlay, anchor = south west, above right = 2cm and 1.4cm of fig2.south west] (overlap) {\includegraphics[width = 2cm]{Figures/frag_size_var_with_seeds_density_inside_q_2_journal_part_1.pdf}};

    \node[overlay, below right=0.6cm and 4cm of fig2.north west] (fig2) {(b)}; 
    \end{tikzpicture}
    \end{subfigure}
\caption{Time traces of average electron density (a) and neon density (b) inside the $q=2$ surface for different mean fragment sizes.}
\label{fig:frag_size_var_den_e_Ne_inside_q_2}
\end{figure}

An overview of the core material assimilation trends can be obtained by plotting the material density inside the $q=2$ surface as shown in \autoref{fig:frag_size_var_den_e_Ne_inside_q_2}. 
The general trend indicates smaller fragments starting to assimilate quicker compared to larger fragments. 
This leads to a quicker increase in the electron and neon density in the plasma .
The trend is a result of smaller fragments having a larger surface area exposed to the background plasma leading to quicker assimilation compared to larger fragments. 

\begin{figure}[htb]
  \setlength{\belowcaptionskip}{-2.01\baselineskip}  
   \centering
     \begin{subfigure}[t]{0.5\textwidth}
         \centering
         \begin{tikzpicture}
            \node (inner sep=0pt) (fig1) at (0,0) {\includegraphics[width=\textwidth]{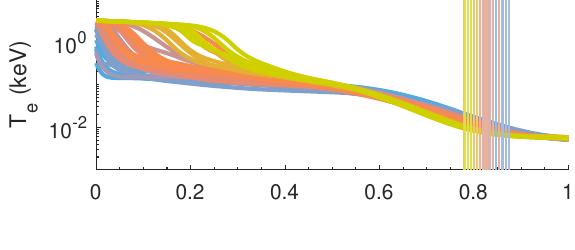}};
            
             \node[overlay, above left=1.4cm and 3 cm of fig1.south east] (fig1) {(a)};
         \end{tikzpicture}         
         \caption{}
         \label{fig:frag_size_var_profiles_at_TQ_onset_a} 
     \end{subfigure}    

          \begin{subfigure}[t]{0.5\textwidth}
         \centering
         \begin{tikzpicture}
            \node (inner sep=0pt) (fig2) at (0,0) {\includegraphics[width=\textwidth]{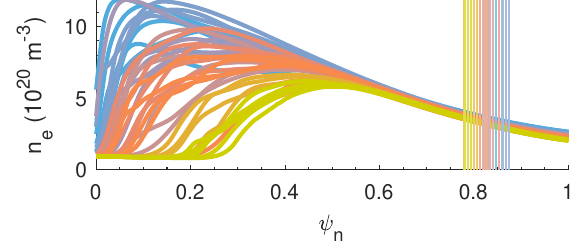}};

        \node[overlay, anchor = south east, above left = 2.3cm and 0
        cm of fig2.south east] (overlap) {\includegraphics[width = 3.8cm]{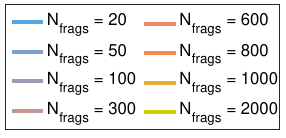}};
        
          \node[overlay, below left=1.3cm and 3 cm of fig2.north east] (fig2) {(b)};          
         \end{tikzpicture}         
         \caption{}
         \label{fig:frag_size_var_profiles_at_TQ_onset_b}
     \end{subfigure}    
\setlength{\belowcaptionskip}{\baselineskip}
\caption{Electron temperature profiles (a) and density profiles (b) at the expected GRE onset for different mean fragment sizes.}
\label{fig:frag_size_var_profiles_at_TQ_onset}
\end{figure}

Assuming the GRE onset condition discussed in \autoref{ssec:characteristic_response}, the time of the GRE onset for each simulation is marked with filled circles in \autoref{fig:frag_size_var_den_e_Ne_inside_q_2}. 
In contrast to the trend discussed above, the larger fragments end up assimilating more material before the GRE onset and the reversal in trend appears due to the location of the deposited material. 
The plasma profiles at the time of expected GRE onset for each simulation are shown in \autoref{fig:frag_size_var_profiles_at_TQ_onset}. 
Due to lower ablation at the edge from larger fragments, the edge cools slowly compared to injections with smaller fragments. 
As a result, larger fragments enable higher core assimilation and at the same time allow for a longer pre-GRE duration compared to smaller fragments. 
Therefore, higher core density and lower core temperatures are observed for larger fragments compared to smaller fragments in \autoref{fig:frag_size_var_profiles_at_TQ_onset}. 

\begin{figure}[H]
\centering
\includegraphics[width=0.5\textwidth]{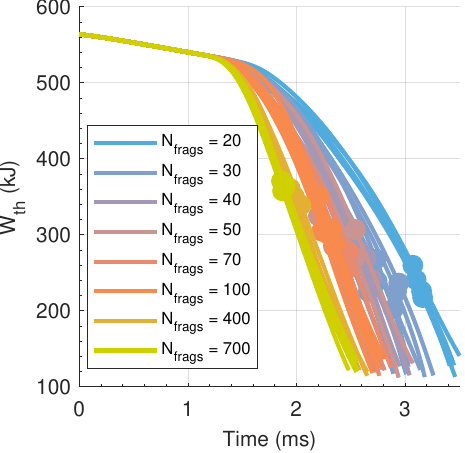}
\caption{Time traces of the plasma thermal energy for different fragment sizes.}
\label{fig:frag_size_var_thermal_energy_time_traces}
\end{figure}

The deeper core penetration and higher radiation from the core also lead to lower remaining plasma thermal energy at the GRE onset as shown in \autoref{fig:frag_size_var_thermal_energy_time_traces}. 
Similar trends in fragment sizes have also been observed in DIII-D INDEX simulations~\cite{lvovskiy_density_2024}. 
Furthermore, qualitative trends of the neon assimilation in the experiments were also studied by Jachmich~\cite{s_jachmich_shattered_2023} by looking at the duration for the plasma current to quench from 100\% to 80\% during the CQ phase $(\tau_{\text{CQ},100\% \rightarrow 80\%})$. 
The duration is an indication of the assimilated neon as the CQ rate is determined by a balance between the external heating and the impurity radiative cooling \cite{wesson_tokamaks_2011}. 
It was reported that larger fragments had a shorter $(\tau_{\text{CQ},100\% \rightarrow 80\%})$, indicating higher neon assimilation, however the trend was more apparent for lower neon content injections.   

\subsubsection{Fragment speeds}
\label{sssec:frag_speed_var_D_Ne}
\begin{figure}[htb]
  \setlength{\belowcaptionskip}{-2.01\baselineskip}  
   \centering
     \begin{subfigure}[t]{0.5\textwidth}
         \centering
         \begin{tikzpicture}
            \node[minimum width=\textwidth, minimum height=0cm] (fig1) {\includegraphics[width=\textwidth]{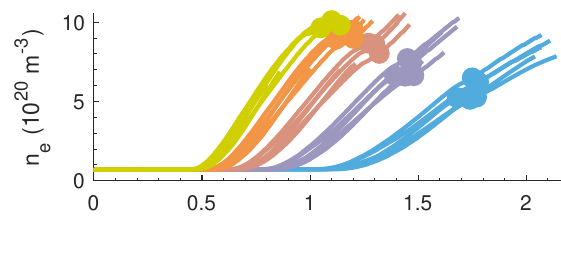}}; 
                    
            \node[overlay, below right=1.8cm and 1.4cm of fig1.north west] (fig1) {(a)};
            
         \end{tikzpicture}         
         \caption{}
         \label{fig:frag_speed_var_den_inside_q_2_a} 
     \end{subfigure}    

          \begin{subfigure}[t]{0.5\textwidth}
         \centering
         \begin{tikzpicture}
                \node[minimum width=0.5\textwidth, minimum height=0cm] (fig2)
                {\includegraphics[width=\textwidth]{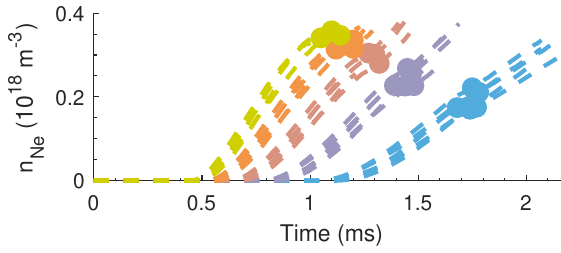}};       

        \node[overlay, anchor = north west, below right = -1.2cm and 1.4cm of fig2.north west] (overlap) {\includegraphics[width = 2.5cm]{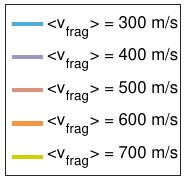}};

        \node[overlay, below right=1.8cm and 1.4cm of fig2.north west] (fig2) {(b)};     
  
    \end{tikzpicture}         
    \caption{}
    \label{fig:frag_speed_var_den_inside_q_2_b}
    \end{subfigure}    
\setlength{\belowcaptionskip}{\baselineskip}
\caption{Time traces of average electron density (a) and neon density (b) inside the $q=2$ surface for different mean fragment speeds.}
\label{fig:frag_speed_var_den_inside_q_2}
\end{figure}

The parametric scan for varying fragment speeds was carried out for 10\% neon and 90\% deuterium pellets assuming a range of mean fragment velocities from $300 - 700$~\si{m/s} with a fixed velocity dispersion. 
An identical fragment size distribution corresponding to 200 sampled fragments was utilised for all the simulations to exclude the effect of varying fragment sizes. 
The resulting time evolution of electron and neon density inside the $q=2$ surface is shown in \autoref{fig:frag_speed_var_den_inside_q_2}. 

\begin{figure}[H]
\centering
\begin{tikzpicture}

    \node (fig1) {\includegraphics[width=0.5\textwidth]{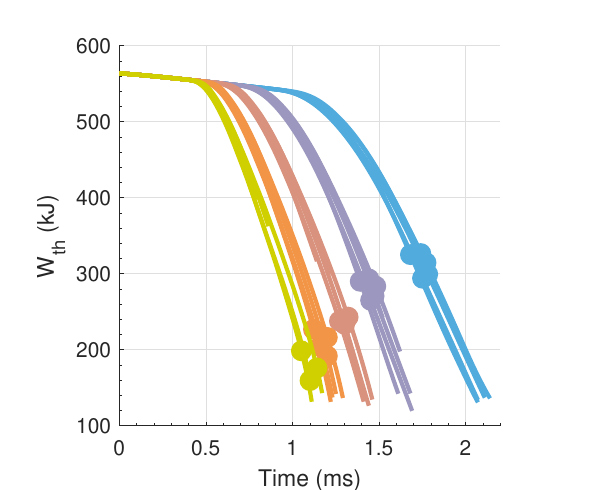}}; 

  \node[overlay, anchor = north east, below left = 0.5cm and 0.5cm of fig1.north east] (overlap) {\includegraphics[width = 2.4cm]{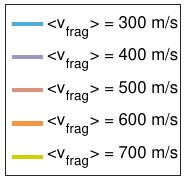}};
\end{tikzpicture}
\caption{Time traces of the plasma thermal energy for different fragment speeds.}
\label{fig:frag_speed_var_thermal_energy}
\end{figure}

\begin{figure}
\centering
\begin{subfigure}[t]{0.5\textwidth}
\begin{tikzpicture}
    \node[minimum width=\textwidth, minimum height=0cm] (fig1)
    {\includegraphics[width=\textwidth]{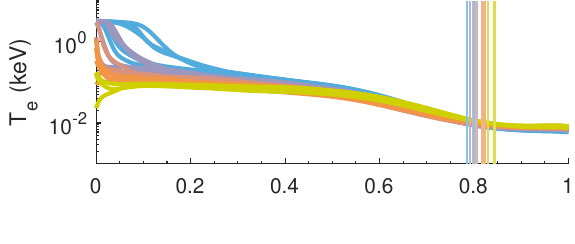}}; 

    \node[above left=1.2cm and 4cm of fig1.south east] (fig1) {(a)};
\end{tikzpicture}
\end{subfigure}

\begin{subfigure}[t]{0.5\textwidth}
\begin{tikzpicture}
  
  \node[minimum width=\textwidth, minimum height=0cm] (fig2)
    {\includegraphics[width=\textwidth]{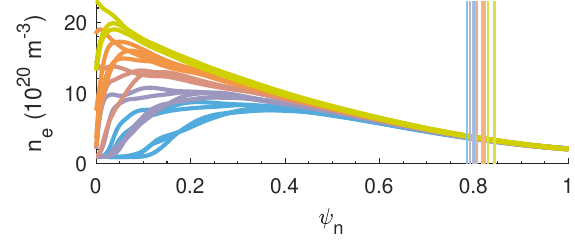}}; 

  \node[above left = 1.2cm and 4cm of fig2.south east] (fig2) {(b)}; 

  \node[overlay, anchor = south east, above left = 0.5cm and -4cm of fig2.south east] {\includegraphics[width = 2.6cm]{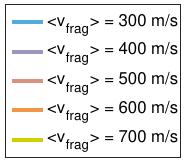}};
\end{tikzpicture}
\end{subfigure}
\caption{Electron temperature profile (a) and density profile (b) at the expected GRE onset for different mean fragment speeds. The vertical lines mark the location of the $q=2$ surface.}
\label{fig:frag_speed_var_Te_ne_profiles_at_TQ}
\end{figure}

The simulations show that faster fragments start to assimilate earlier than slower fragments due to quicker penetration of the injected material. 
Despite the faster fragments also having a shorter pre-GRE duration, more material is assimilated before the cold front reaches the $q=2$ surface.
This effect can be explained by the injected material being able to penetrate deeper and assimilate more material towards the core before the formation and propagation of the cold front. 
The resultant electron density and temperature profiles at the moment the cold front reaches the $q=2$ surface are shown in \autoref{fig:frag_speed_var_Te_ne_profiles_at_TQ}. 
The profiles discernibly indicate deeper assimilation of material injected with faster fragments, leading to stronger radiative cooling of the plasma core and consequently, lower core temperatures. 
Time evolution of the plasma thermal energy is shown in \autoref{fig:frag_speed_var_thermal_energy}. 
The quicker and higher assimilation of the injected material for faster fragments also leads to more thermal energy being radiated away in the pre-GRE phase.

Similar trends have been observed for INDEX simulations for DIII-D~\cite{lvovskiy_density_2024} and ITER~\cite{matsuyama_transport_2022} which show faster fragments leading to higher assimilation. Experimental measurements of $\tau_{\text{CQ},100\% \rightarrow 80\%}$ in AUG~\cite{s_jachmich_shattered_2023} also indicate faster fragments leading to higher neon assimilation. 
An aspect of the SPI system not considered in the above simulations is the effect of the gas generated during the shattering process. 
Gebhart et\mbox{.} al\mbox{.}~\cite{Gebhart2020Experimental} have studied the amount of gas produced during the shattering process and found it to have a strong dependence on $\chi_R$, which is the ratio of pellet kinetic energy to the threshold kinetic energy. 
At higher values of $\chi_R$, $<30\%$ of the pellet mass has been detected from the measured fragment size distributions from fast cameras. 

\subsubsection{Pellet composition scan}
Finally, we discuss a set of simulations for varying neon composition of the pellet. 
The neon fraction in the pellet was varied from 0.1\% to 50\% of a 10~mm long and 8~mm diameter pellet, which corresponds to $\sim 1.2{\cdot}10^{19}$ to $8.9{\cdot}10^{21}$ atoms. 
To eliminate the influence of fragment parameters, the same fragment size and speed distribution was utilised in all the simulations. 
200 fragments were sampled with a mean fragment velocity of $230~\si{m/s}$.

\begin{figure}
\centering
\begin{subfigure}[t]{0.5\textwidth}
    \centering
    \begin{tikzpicture}
    \node[minimum width=\textwidth, minimum height=0cm] (fig1)
    {\includegraphics[width=\textwidth]{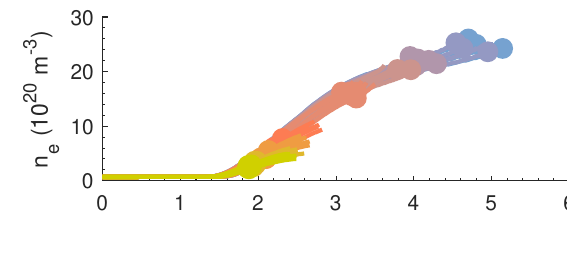}}; 
    \node[overlay, below right=1cm and 2cm of fig1.north west] {(a)};
  
    \end{tikzpicture}
\end{subfigure}

\begin{subfigure}[t]{0.5\textwidth}
\centering
\begin{tikzpicture}
    \node[minimum width=\textwidth, minimum height=0cm] (fig2) {\includegraphics[width=\textwidth]{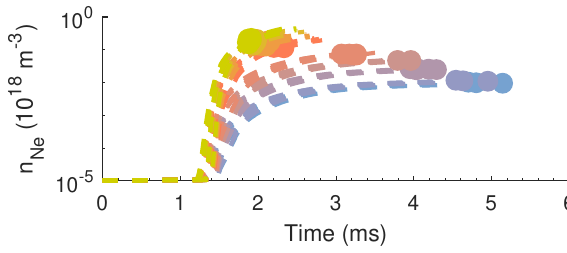}}; 
    
  \node[overlay, below right=1cm and 2cm of fig2.north west] (fig2) {(b)}; 

  \node[overlay, anchor = south east, above right = 1.2cm and 1.6cm of fig2.south east]{\includegraphics[width = 3.8cm]{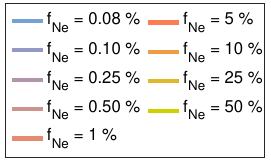}};  
  
  \end{tikzpicture}
  \end{subfigure}
\caption{Time traces of average electron density (a) and neon density (b) inside the $q=2$ surface for different pellet composition.}
\label{fig:xmol_var_e_Ne_density}
\end{figure}

The time evolution of electron and neon density inside the $q=2$ surface is shown in \autoref{fig:xmol_var_e_Ne_density}. Injecting higher amounts of neon lead to higher neon density rise and lower electron density rise. 
The neon assimilation varies strongly for trace amounts ($< 5\%$) of injected neon. At higher neon quantities, the neon assimilation changes minimally indicating a self-regulating balance between decrease in plasma temperature due to stronger radiative cooling which in turn reduces further ablation. 
Similar to the previous scans, the circular scatter points in \autoref{fig:xmol_var_e_Ne_density} indicate the expected time of the GRE onset. 
Following the trend of neon assimilation, the pre-GRE duration decreases drastically with increasing amounts of injected neon for trace amounts ($< 5\%$) and only minimally for larger neon amounts.

\begin{figure}[H]
\centering
\begin{subfigure}[t]{0.5\textwidth}
\begin{tikzpicture}
    \node[minimum width=\textwidth, minimum height=0cm] (fig1) {\includegraphics[width=\textwidth]{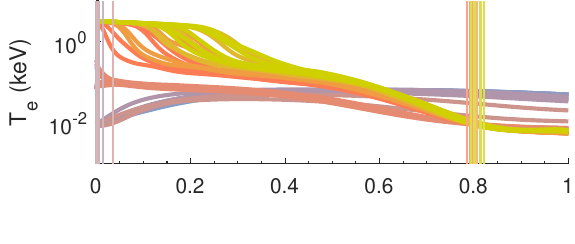}}; 

    \node[below left=0cm and 4cm of fig1.north east] (fig1) {(a)};
\end{tikzpicture}        
\end{subfigure}

\begin{subfigure}[t]{0.5\textwidth}
\centering
\begin{tikzpicture}                
    \node[minimum width=\textwidth, minimum height=0cm] (fig2) {\includegraphics[width=\textwidth]{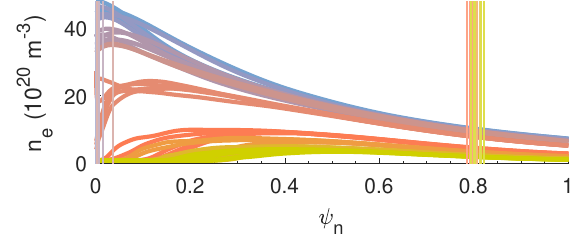}}; 

  \node[overlay, anchor = north west, above left = -1.5cm and 0.2cm of fig2.north east]{\includegraphics[width = 3.4cm]{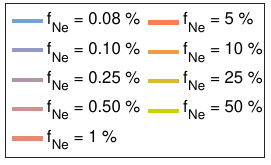}};

  \node[below left=0cm and 4cm of fig2.north east] (fig2) {(b)}; 

  \end{tikzpicture}
  \end{subfigure}
\caption{Electron temperature profile (a) and density profile (b) at the GRE onset for different pellet composition. The vertical lines mark the location of the $q=2$ surface.}
\label{fig:xmol_var_profiles_at_TQ}
\end{figure}

\begin{figure}[H]
\centering    
\begin{tikzpicture}
    \node (fig1) {\includegraphics[width=0.5\textwidth]{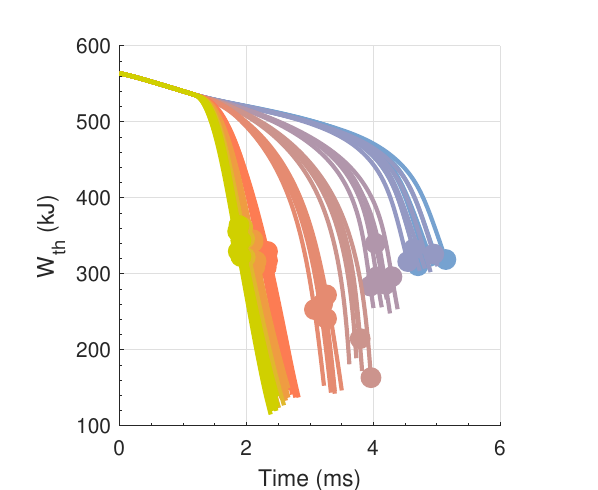}}; 

  \node[overlay, anchor = north east, below left = 0cm and 0cm of fig1.north east] (overlap) {\includegraphics[width = 3.5cm]{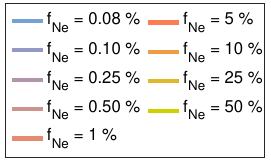}};
\end{tikzpicture}
\caption{Time traces of the plasma thermal energy for different pellet composition.}
\label{fig:xmol_var_thermal_energy}
\end{figure}

The electron density and temperature profiles at the time of expected GRE onset are shown in \autoref{fig:xmol_var_profiles_at_TQ}. 
While the larger neon amounts ($>5\%$) show the formation of a cold front at the plasma edge before the plasma core cools down, injections with lower amounts or neon ($<1\%$), lead to an inside-out temperature collapse. 
For these low neon amount injections, the fragments have enough time to travel to and deposit more material in the hot plasma core because of the higher ablation rates, before a cold-front forms at the plasma edge. 
In such cases, the MHD process leading to a disruption might be drastically different compared to the outside-in temperature collapse. Nevertheless, for the sake of comparison, the same GRE onset condition is used for these cases.
The inside-out temperature collapse also allows for more thermal energy to be radiated away before the expected GRE event as indicated in \autoref{fig:xmol_var_thermal_energy} by the circular scatter points. 
While the time evolution of the thermal energy follow the general trend of quicker radiation with increasing neon content, a non-monotonic trend of the remaining thermal energy at GRE-onset is observed. However, this trend requires further investigation as it may be influenced by the nature of the MHD activity that leads to a disruption in such cases.

A limitation of the present scan is that it does not take into account the plasmoid drift~\cite{matsuyama_enhanced_2022} which might shift the ablated material towards the magnetic low-field-side (LFS) especially for lower neon content injections and pure D injections. 
While it has been shown that the presence of 2-5\% neon can substantially suppress the plasmoid drift~\cite{matsuyama_enhanced_2022,kong_interpretative_2024}, a more comprehensive relation between neon content and the extent of drift suppression is still under investigation. 


\begin{figure}[H]
    \centering
\begin{tikzpicture}

    \node (fig1) {\includegraphics[width=0.5\textwidth]{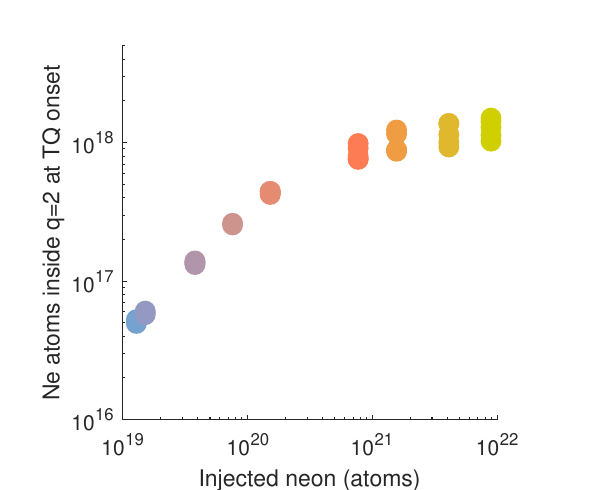}};   

  \node[overlay, anchor = south east, above left = 1.0cm and 1.4cm of fig1.south east] {\includegraphics[width = 1.5cm]{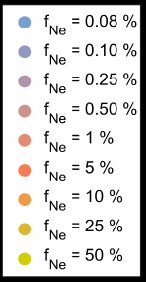}};
\end{tikzpicture}
\caption{Assimilated neon inside the $q=2$ surface at the GRE onset for varying neon content injections.}
\label{fig:xmol_var_Ne_inside_q_2_at_TQ_onset}
\end{figure}

Experimental indications of neon assimilation in the experiments have been indirectly provided by measuring the radiated energy fraction throughout the disruption~\cite{heinrich_analysis_2024} and the value of $\tau_{\text{CQ},100\% \rightarrow 80\%}$~\cite{s_jachmich_shattered_2023}. 
Both the quantities have a strong dependence on the neon content for lower neon content values: $10^{19}$ to $10^{21}$ atoms or 0 to 5\% of 8 \si{mm} diameter and 10 \si{mm} length pellets. 
While for neon content $>10^{21}$ atoms, both quantities have minimal changes in their values indicating a self-regulating cooling effect of the injected neon as mentioned above. 
The trend in neon assimilation, as observed in the present pre-GRE simulations, provide an indirect comparison to the experimental indications of the neon assimilation discussed above.  
\autoref{fig:xmol_var_Ne_inside_q_2_at_TQ_onset} shows the neon assimilation inside the $q=2$ surface at the GRE onset, clearly indicating the strong dependence on neon content initially and saturation for larger neon quantities. 

\subsection{General trends}
\begin{figure}[htb]
\setlength{\belowcaptionskip}{-2.01\baselineskip}  
\begin{center}
\begin{subfigure}[t]{0.49\textwidth}
\begin{tikzpicture}
    \node[minimum width=\textwidth, minimum height=0cm] (fig1) {\includegraphics[width=\textwidth]{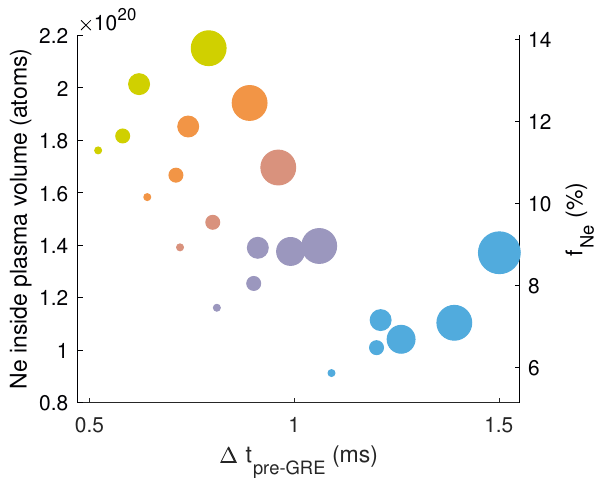}}; 

  \node[overlay, below left = 0.8cm and 1.5cm of image] at (fig1.north east) {(a: 10\% Ne)};
  
\end{tikzpicture}
\label{}
\caption{}
\end{subfigure}
\hfill
\begin{subfigure}[t]{0.49\textwidth}
\begin{tikzpicture}
    \node[minimum width=\textwidth, minimum height=0cm] (fig2) {\includegraphics[width=\textwidth]{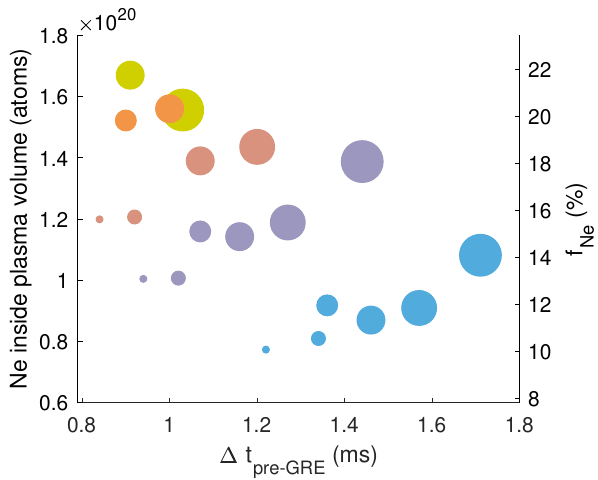}};     

      \node[overlay, below left = 0.8cm and 1.5cm of image] at (fig2.north east) {(b: 5\% Ne)};
    
    \end{tikzpicture}
    \label{}
    \caption{}
    \end{subfigure}
\end{center}
    
\begin{subfigure}[t]{0.5\textwidth}
\begin{tikzpicture}
\node[minimum width=\textwidth, minimum height=0cm] (fig3) {\includegraphics[width=\textwidth]{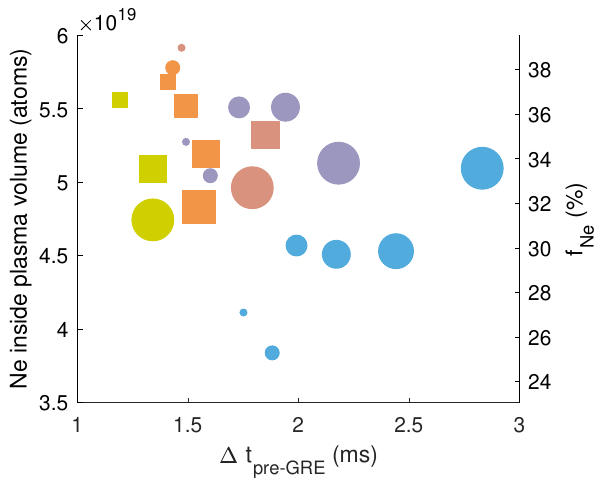}}; 

\node[overlay, below left = 0.8cm and 1.5cm of image] at (fig3.north east) {(c: 1\% Ne)};
\end{tikzpicture}
\caption{}
\label{fig:pre_TQ_vs_den_2D_scans_c}
\end{subfigure}
\hfill
\begin{subfigure}[t]{0.5\textwidth}
\begin{tikzpicture}
  \node[overlay, minimum width=0.5\textwidth, minimum height=0cm] at (5,4)  (leg1) {\includegraphics[width = 3.5cm]{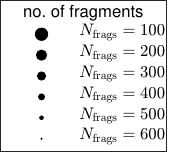}};
  
  \node[overlay, minimum width=0.5\textwidth, minimum height=0cm] at (2,3) (leg2) {\includegraphics[width = 3.7cm]{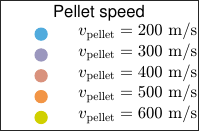}};
\end{tikzpicture}
\end{subfigure}
\setlength{\belowcaptionskip}{\baselineskip}
\caption{Assimilated neon inside plasma volume and pre-GRE duration for varying fragment sizes and speeds in AUG H-mode SPI discharges. The left axes show absolute neon assimilation while the right axis shows the fraction of injected neon that is assimilated. Different colors indicate different mean fragment speeds while the different marker sizes correspond to different mean fragment size with the largest markers corresponding to largest sizes and vice versa. The square markers in part (c) correspond to simulations which had an inside-out temperature collapse.}
\label{fig:pre_TQ_vs_den_2D_scans}
\end{figure}
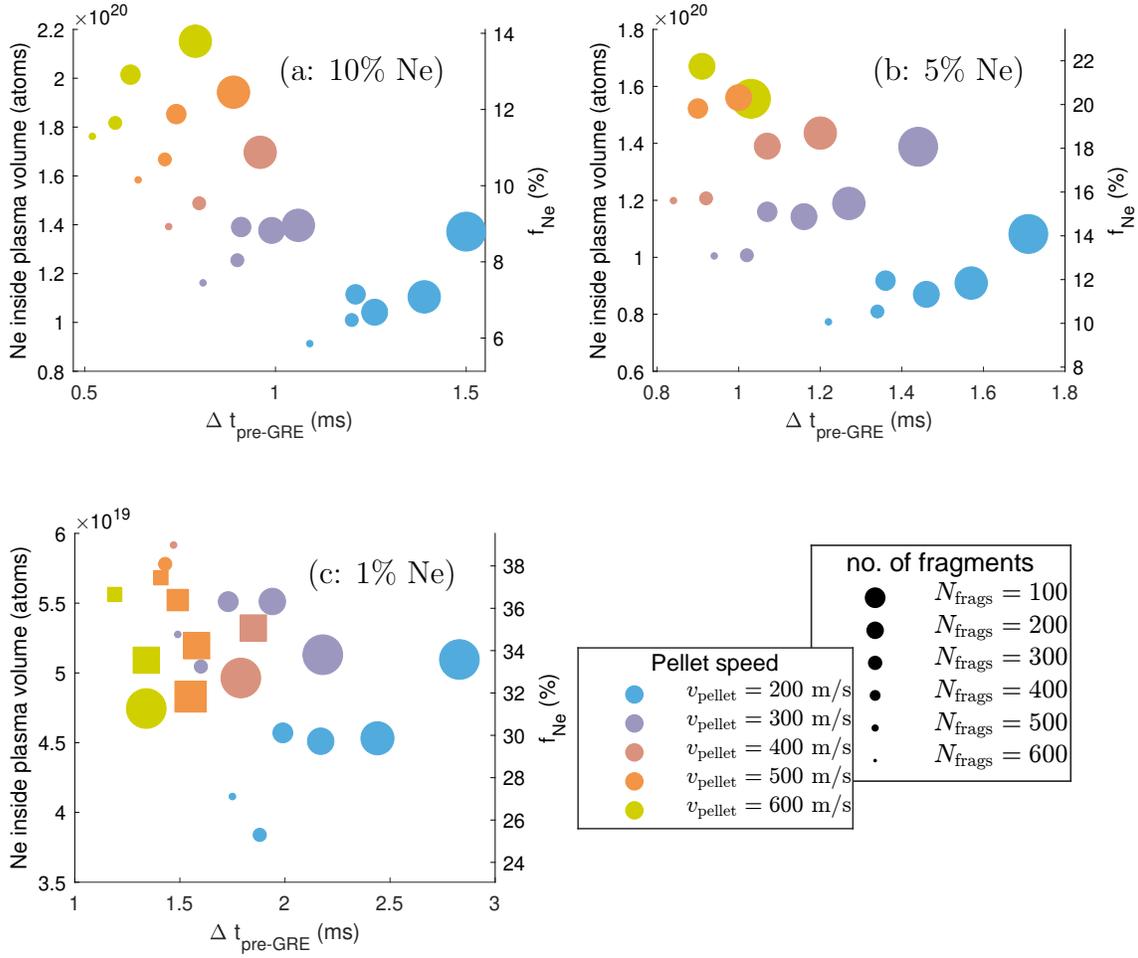

The above scans are instructive in understanding the plasma response to different fragment parameters. More general parameter trends are reported in this section. 
These simulations span the expected range of fragment size and speed parameters expected after the shattering for AUG SPI~\cite{peherstorfer_fragmentation_2022}. 
The expected pre-GRE time and the assimilated neon fraction in the plasma volume are shown in \autoref{fig:pre_TQ_vs_den_2D_scans} for variations in all three parameters. Overall, larger fragments allow for higher assimilation and a longer pre-GRE duration compared to smaller fragments. 
While for fragment speeds, a general trade-off between higher material assimilation and shorter pre-GRE duration is observed. 
Comparing the trends across different neon composition, higher neon content allow for higher absolute neon assimilation however the fraction of assimilated neon is lower. 

The square markers in \autoref{fig:pre_TQ_vs_den_2D_scans_c} correspond to simulations which had an inside-out temperature collapse. 
In this case, the GRE-onset condition might be modified and hence the trends might not be valid but are shown for the sake of comparison. The dependence of material assimilation on fragment sizes and speeds seems to hold across pellets with different neon content at least for the simulations where an edge temperature collapse is observed. Further investigation is required for injections that lead to an inside-out temperature collapse. 

\subsection{Pure deuterium pellets}
\label{ssec:pure_deuterium_pellets}
Pure hydrogen pellets have also been proposed for the ITER DMS~\cite{nardon_possible_2020} in a staggered injection scheme, where the first injection will increase the plasma density for RE avoidance, which will then be followed by a mixed hydrogen-neon injection for heat load mitigation. 
However, pure hydrogen and deuterium injections (used in AUG) experience a plasmoid drift that can limit that limit the material assimilation to the plasma edge~\cite{matsuyama_enhanced_2022, vallhagen_drift_2023, muller_high_2002}. 
Hence, to simulate pure \si{D_2} injections, we have simulated the radial drift of the ablated material with a back-averaging model in INDEX~\cite{lvovskiy_density_2024}. 
The back-averaged density rise from a single fragment is estimated as 
\begin{equation}
\Delta n_\text{dep}=\frac{\Delta N}{ (1+\backavg) V_{\text{p}}(\rho = 1)-V_{\text{p}}\left(\rho_\text{f}\right)},
\end{equation}
where $\Delta N$ is the number of ablated atoms and $V_\text{p}$ is the plasma volume and $\rho_\text{f}$ is the location of the fragment. 
The extent of the drift is governed by the parameter $\backavg$. 
While more consistent models of the plasmoid drift for shattered pellets, similar to non-shattered pellets~\cite{kochl_pellet_2008}, can provide a value for $\backavg$, presently interpretive simulations need to be used to match the simulated material assimilation with experimental plasma density measurements, as previously done with INDEX DIII-D simulations~\cite{lvovskiy_density_2024} and JOREK JET simulations~\cite{kong_interpretative_2024}.
We use Thomson scattering~(TS) measurements for AUG SPI discharges to determine $\backavg$, and the simulation results are then compared to the line-integrated density measurements.
The poloidal locations of these diagnostics in the AUG tokamak are shown in \autoref{fig:AUG_INDEX_inputs_poloidal_cross_section}. The Thomson scattering diagnostic has the same poloidal location as the core channel of the interferometer.

\begin{figure}[htb]
\centering
\includegraphics[width = 15cm]{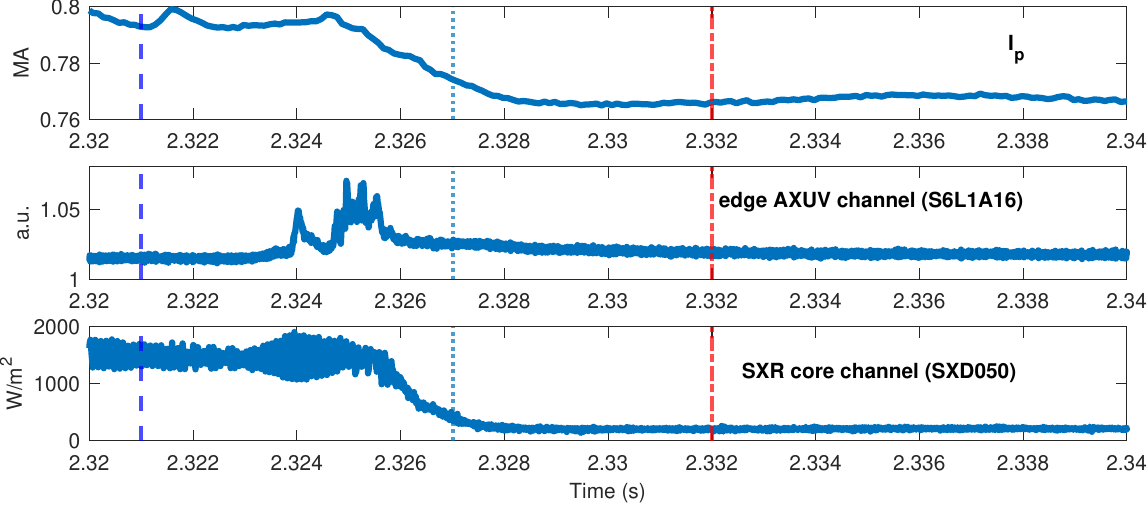}
\caption{Experimental plasma parameters of \#40743 during a pure \si{D_2} SPI entering the plasma at $\approx 2.324 \si{ms}$. Three vertical dashed lines mark timestamps whose corresponding density profiles are shown in \autoref{fig:40743_TS_measurements}.}
\label{fig:40743_params}
\end{figure}

AUG \#40743 is used as a representative discharge to determine the value of $\backavg$ and the SPI related plasma parameters are shown in \autoref{fig:40743_params} with the relevant line of sights shown in \autoref{fig:AUG_INDEX_inputs_poloidal_cross_section}. 
The discharge had a dual injection with the first pure \si{D_2} injection entering the plasma around 2.324~seconds as detected from the sudden rise in the edge AXUV \cite{bernert_application_2014} channel. 
The first injection leads to a minor drop in the plasma current, but does not lead to a disruption. 
Due to the dilution cooling from the injected deuterium, the plasma temperature drops and the soft X-ray (SXR) signal from one of the core channels decreases as its sensitivity reduces drastically below 1~keV~\cite{igochine_hotlink_2010}. 
The second injection (not shown) occurs at $\sim 30 \si{ms}$ after the first injection, around $2.355\si{ms}$.

\begin{figure}[htb]
\centering
\includegraphics[width=0.5\textwidth]{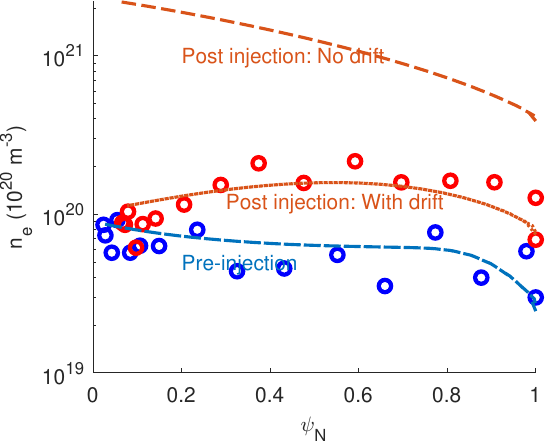}
\caption{Comparison of experimental and simulated electron density profiles considering the effect of the plasmoid drift.}
\label{fig:40743_TS_with_without_back_avg} 
\end{figure}

\begin{figure}[htb]
\centering
\includegraphics[width = 0.5\textwidth]{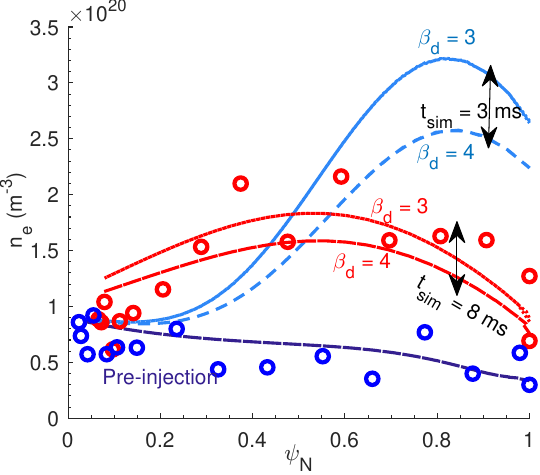}
\caption{Comparison of experimental and simulated electron density profiles for $\backavg = 3$ and $\backavg = 4$.}
\label{fig:40743_TS_measurements} 
\end{figure}

Vertical lines in \autoref{fig:40743_params} mark the time stamps of the TS profiles, which are plotted against normalised poloidal flux in \autoref{fig:40743_TS_with_without_back_avg} and \autoref{fig:40743_TS_measurements}.
To compare the effect of the back-averaging process on the material assimilation, we compare post-injection synthetic TS profiles with the experimental measurements shown in \autoref{fig:40743_TS_with_without_back_avg}. 
Comparison between the simulated and experimental density measurements shows that without the inclusion of plasmoid drift, the material assimilation is drastically overestimated. 

Considering the effect of the plasmoid drift, it is instructive to study the plasma dynamics during pure \si{D_2} injections. 
An additional time slice marked with a (light-blue) dashed line in \autoref{fig:40743_params} indicates the time when the maximum volume-averaged plasma density is reached in the simulations with the plasmoid drift.
At this time, most of the ablation has taken place and the rest of the material has been shifted out of the plasma by the back-averaging process. 
The resultant density profile at this point is shown in  in \autoref{fig:40743_TS_measurements} marked by $t_\text{sim} = 3$.
Beyond this point, the plasma temperature and density profiles evolve due to heat and particle diffusive transport. 
Simulated density and temperature profiles from the synthetic TS diagnostic for two $\backavg$ values are compared with the experimental measurements in \autoref{fig:40743_TS_measurements}. 
The results indicated that $\backavg = 4$ has the best match with the experimental measurements over the entire plasma domain. 
Hence, this value of $\backavg$ is utilised for all pure \si{D_2} injections in this paper.
We also assume, for simplicity, that the value of $\backavg$ remains constant regardless of different fragment sizes and speeds. 


\begin{figure}[htb]
\centering
\includegraphics[width = 0.5\textwidth]{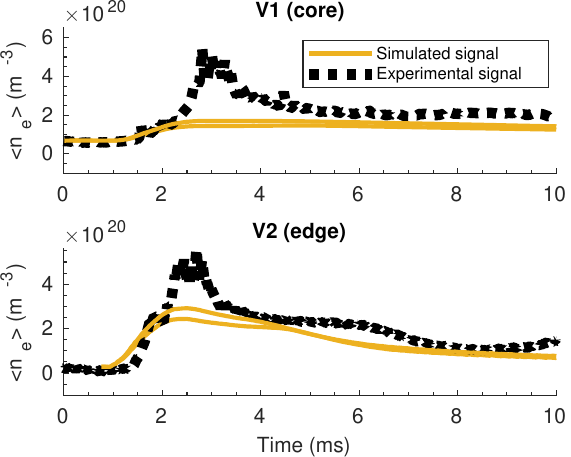}
\caption{Comparison of experimental (black dotted) and simulated (yellow solid) line-averaged density along the \si{CO_2} interferometer line of sights for \#40743. Simulations with $\backavg = 3$ lead to a slightly higher interferometer signal compared to $\backavg = 4$ due to a slightly higher density increase as also shown in \autoref{fig:40743_TS_measurements}.}
\label{fig:40743_interf_measurements} 
\end{figure}

To further justify the value of $\backavg$, we also carried out comparisons with the experimental measurements of the line-integrated density from the \si{CO_2} laser interferometer at AUG~\cite{mlynek_infrared_2012,mlynek_simple_2017}, as shown in \autoref{fig:40743_interf_measurements}. 
The comparison is carried out for two channels, referred to as a core and an edge channel whose poloidal locations are marked in \autoref{fig:AUG_INDEX_inputs_poloidal_cross_section}. 
Given that the line-integrated measurements can be sensitive to the plasma movement, which are not taken into account in the present experimental data, the value of $\backavg = 4$ is preferred due to better matching with the TS measurements.

Another feature to note in the comparison is that the peak density rise in the two channels, at $\sim 3\si{ms}$, is not captured by the synthetic measurements. 
A possible explanation for this discrepancy is the nature of the line-integrated measurements in the experiments, which can detect the plasmoid passing through its line of sight and registering a large increase in the electron density which has not been accounted for in the INDEX synthetic diagnostic. 
As the plasmoid(s) eventually assimilate into the background plasma, the experimental signal decays to a stable value which is comparable to the synthetic diagnostic output from INDEX.  
It should be noted that the present simulations do not take into account the presence of intrinsic impurities in the plasma which can affect the plasma dynamics, especially for pure D injections, depending on the quantity~\cite{kong_interpretative_2024, nardon_fast_2020}.

\subsubsection{Fragment sizes \label{sssec:frag_size_var_D}} 

\begin{figure}[htb]
  \setlength{\belowcaptionskip}{-2.01\baselineskip}  
   \centering
     \begin{subfigure}[t]{0.5\textwidth}
         \centering
         \begin{tikzpicture}
            \node (inner sep=0pt) (image) at (0,0) {\includegraphics[width=\textwidth]{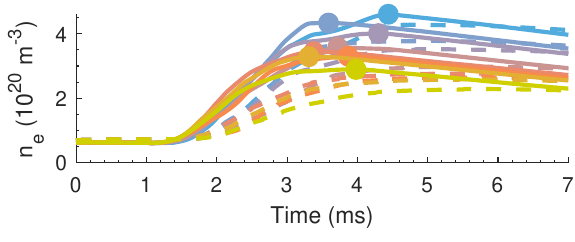}};

        \node[overlay, anchor = south east, above left = 0cm and -1.1
        cm of image.south east] (overlap) {\includegraphics[width = 3.8cm]{Figures/Deuterium_size_var_e_density_time_traces_and_profiles_at_max_vol_density_journal_part_1.pdf}};
            
            \node[overlay, below right = 0.7cm and 1.4cm of image] at (image.north west) {(a)};
         \end{tikzpicture}         
         \caption{}
         \label{fig:size_var_pure_D_den_inside_q2_vol_top} 
     \end{subfigure}    

          \begin{subfigure}[t]{0.5\textwidth}
         \centering
         \begin{tikzpicture}
            \node (inner sep=0pt) (image) at (0,0) {\includegraphics[width=\textwidth]{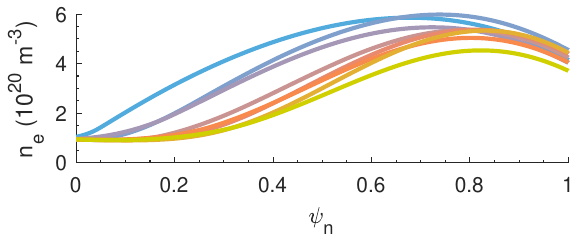}};        
        
        \node[overlay, below right = 0.7cm and 1.4cm of image] at (image.north west) {(b)};
        
         \end{tikzpicture}         
         \caption{}
         \label{fig:size_var_pure_D_den_inside_q2_vol_bottom} 
     \end{subfigure}    
    
\setlength{\belowcaptionskip}{\baselineskip}
\caption{(a) Time traces of the electron density averaged inside the plasma volume in solid lines and averaged inside the $q=2$ surface in dashed lines. (b) Electron density profiles at the time of maximum volume averaged density for each simulation as marked by circular scatter points in the top figure.}
\label{fig:size_var_pure_D_den_inside_q2_vol} 
\end{figure}

After obtaining an estimate of $\backavg$, we carried out parametric scans for pure \si{D_2} SPI. Similar to \autoref{sssec:frag_size_var_D_Ne}, fragment sizes are varied by changing the number of sampled fragments from a fixed pellet volume and sampling from a distribution with the same mean fragment velocity. 
To study the material assimilation, time traces of the electron density averaged inside plasma volume and inside the $q=2$ surface for simulations with different fragment sizes are shown in \autoref{fig:size_var_pure_D_den_inside_q2_vol_top}. 
Larger fragments seem to lead to higher material assimilation throughout most of the post-injection phase.

Due to the higher edge material assimilation resulting from the plasmoid drift, the volume-averaged density (solid lines) initially remains higher than the average electron density inside $q=2$ surface (dashed lines). 
Eventually, as the plasma density evolves due to the diffusive transport, a crossover happens and volume-averaged electron density becomes higher than the average density inside the $q=2$ surface for all the simulations.
As majority of the ablation is concluded by the time the maximum volume-averaged density is achieved, electron density and temperature profiles at this time are plotted in \autoref{fig:size_var_pure_D_den_inside_q2_vol_bottom} for all simulations.

Larger fragments enable initial deeper penetration into higher temperature regions resulting in higher material assimilation compared to smaller fragments, before the rest of the material is shifted out of the plasma. 
In all the pure \si{D_2} simulations, there is no cold front formation due to the lack of radiative cooling associated with neon atoms.

\subsubsection{Fragment speeds}
\begin{figure}[H]
  \setlength{\belowcaptionskip}{-2.01\baselineskip}  
   \centering
     \begin{subfigure}[t]{0.5\textwidth}
         \centering
         \begin{tikzpicture}
            \node (inner sep=0pt) (fig1) at (0,0) {\includegraphics[width=\textwidth]{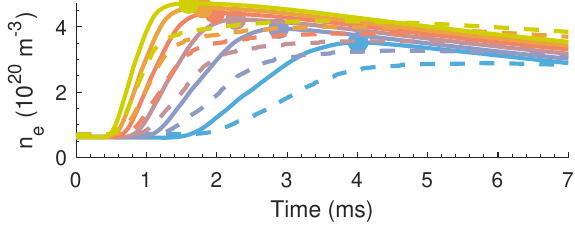}};    
            
            \node[overlay, below left = 1.2cm and 2cm of image] at (fig1.north east) {(a)};
            
         \end{tikzpicture}         
         \caption{}
         \label{fig:speed_var_pure_D_den_inside_q2_vol_and_profiles_a} 
     \end{subfigure}    

          \begin{subfigure}[t]{0.5\textwidth}
         \centering
         \begin{tikzpicture}
            \node (inner sep=0pt) (fig2) at (0,0) {\includegraphics[width=\textwidth]{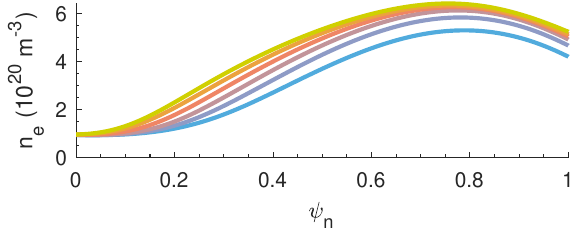}};    

        \node[overlay, anchor = south east, above right = 2cm and 1.1
        cm of fig2.south west] (overlap) {\includegraphics[width = 2.1cm]{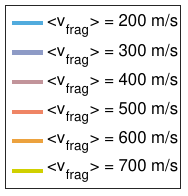}};

        \node[overlay, below left = 0.8cm and 2cm of image] at (fig2.north east) {(b)};
        
         \end{tikzpicture}         
         \caption{}
         \label{fig:speed_var_pure_D_den_inside_q2_vol_and_profiles_b} 
     \end{subfigure}    
    
\setlength{\belowcaptionskip}{\baselineskip}
\caption{(a) Time traces of the electron density averaged inside the plasma volume in solid lines and averaged inside the $q=2$ surface in dashed lines. (b) Electron density profiles at the time of maximum volume averaged density for each simulation as marked by circular scatter points in the top figure.}
\label{fig:speed_var_pure_D_den_inside_q2_vol_and_profiles} 
\end{figure}

The effect of varying fragment speeds for pure \si{D_2} injections is studied similarly as done in \autoref{sssec:frag_speed_var_D_Ne} by changing the mean fragment speed and using the same fragment size distribution for all the cases. 
The time traces of average electron density for the set of simulations are shown in \autoref{fig:speed_var_pure_D_den_inside_q2_vol_and_profiles}. 
Similar to the mixed D/Ne injections, these results also indicate faster fragments having overall higher material assimilation throughout the simulation. 
The faster fragments also allow for slightly higher assimilation towards the intermediate plasma region compared to the slower fragments as shown in \autoref{fig:speed_var_pure_D_den_inside_q2_vol_and_profiles_b}. 

Analysis of the line-integrated density measurements in pure deuterium injections in AUG have been carried out by Jachmich~\cite{s_jachmich_shattered_2023} using the peak density rise from the interferometer. 
The experimental trends of pure \si{D_2} pellet assimilation also indicate higher assimilation for larger and faster fragments. 
However, quantitative comparisons require further experimental analysis as the peak density values can be much higher than the final intermediate densities as also shown in \autoref{fig:40743_interf_measurements}.

\section{Summary \label{sec:summary}}

We used the INDEX code to study the material assimilation trends for different SPI fragment parameters in the AUG tokamak. The effect of three parameters: fragment sizes, speeds and pellet composition is reported. Mixed deuterium-neon as well as pure deuterium injections are simulated and qualitative comparisons with the previously reported trends is carried out. 
We use a semi-empirical global reconnection event onset condition to study the material assimilation before the GRE event. 
For mixed deuterium-neon injections, smaller fragments assimilate quicker but also cool down the plasma edge quicker, hereby limiting the assimilation before the GRE onset. 
Larger fragments take longer to assimilate due to the smaller surface area, however can penetrate deeper in the plasma and assimilate more material before the plasma edge cools down. 
As a result, larger fragments end up with higher assimilation before the GRE onset. 
Faster fragments also lead to higher material assimilation albeit with shorter expected pre-GRE duration as the bulk material reaches the plasma quicker for faster fragments.
Nevertheless, the faster fragments ablate and assimilate more towards the intermediate plasma regions compared to slower fragments before the plasma edge cools down.
Lower neon content injections can lead to an inside-out temperature collapse as more neon ends up in the plasma core and cools down the core quicker compared to the edge. 
While the assimilated neon content increases with increasing amounts of injected neon, there is a self-regulating cooling mechanism which limits the variation in neon assimilation to minimal amounts for injections with more than $10^{21}$ injected neon atoms.  

A back-averaging model is used to simulate the plasmoid drift for pure deuterium injections. The back averaging parameter is determined through an interpretive simulation of an experimental pure \si{D_2} injection discharge and its Thomson scattering measurements.
Experimental material assimilation could only be matched by enabling the back-averaging process. 
Parametric scans of fragment parameters for pure \si{D_2} injections also indicate larger and faster fragments leading to higher assimilation. 
The initial material deposition is limited to the plasma edge due to the plasmoid drift and the core density increases on a longer timescale by diffusive transport.
 
The trends are in line with previously reported qualitative experimental trends, i.e. larger and faster fragments leading to higher assimilation for both mixed D/Ne and pure \si{D_2} injections. 
The neon assimilation trends for varying neon composition are also corroborated by experimental trends of radiated energy fraction and CQ duration. 

\paragraph{Acknowledgements}

\noindent
The authors are grateful to Roger Jaspers, Eric Nardon, Di Hu and Peter Halldestam for fruitful discussions.
This work has been carried out within the framework of the EUROfusion Consortium, funded by the European Union via the Euratom Research and Training Programme (Grant Agreement No 101052200 — EUROfusion). Views and opinions expressed are however those of the author(s) only and do not necessarily reflect those of the European Union, the European Commission, or the ITER Organization. Neither the European Union nor the European Commission can be held responsible for them.
This work received funding from the ITER Organization under contract IO/20/IA/43-2200. The ASDEX-Upgrade SPI project has been implemented as part of the ITER DMS Task Force programme. The SPI system and related diagnostics have received funding from the ITER Organization under contracts IO/20/CT/43-2084, IO/20/CT/43-2115, IO/20/CT/43-2116.

\bibliographystyle{iaea_papp_natbib}
\bibliography{INDEX_SPI_2024}

\end{document}